\documentclass[prl,aps,twocolumn,groupedaddress,showpacs,floatfix]{revtex4}
\usepackage{amsmath,amssymb,multirow,epsfig,bm}

\newcommand{\etal}{{\it et al.,\;}}
\newcommand{\beq}{\begin{equation}}
\newcommand{\eeq}{\end{equation}}
\newcommand{\bea}{\begin{eqnarray}}
\newcommand{\eea}{\end{eqnarray}}

\newcommand{\benn}{\begin{displaymath}}
\newcommand{\eenn}{\end{displaymath}}

\begin{document}

\title{Induced P-wave superfluidity within full energy-momentum dependent Eliashberg approximation in asymmetric dilute Fermi gases} 
 
\author{ Aurel Bulgac and Sukjin Yoon}
\affiliation{Department of Physics, University of Washington, Seattle, WA 98195--1560, USA}

\begin{abstract}
We consider a very asymmetric system of Fermions with an interaction characterized by a positive scattering length only. The minority atoms pair and form a BEC of dimers, while the surplus fermions interact only indirectly through the exchange of Bogoliubov sound modes. This interaction has a finite range, the retardation effects are significant and the surplus fermions will form a P-wave superfluid. We compute the P-wave pairing gap in the BCS and Eliashberg with only energy dependence approximations, and demonstrate their inadequacy in comparison with a full treatment of the momentum and energy dependence of the induced interaction. The pairing gap computed with a full momentum and energy dependence is significantly larger in magnitude and that makes it more likely that this new exotic paired phase could be put in evidence in atomic trap experiments.

\end{abstract}

\date{\today}


 
\maketitle
 
\subsection{I. Introduction}
 
It is still largely an unsettled issue what happens to superfluidity in asymmetric fermionic systems. The asymmetry is measured by the ratio of the minority {\it versus} majority number densities 
$x=n_\downarrow/n_\uparrow$. If the only interaction between the fermions is between the two species and if that interaction is attractive, the two extremum cases are clear: a fully paired phase for $x=1$, spanning the full range from a Bardeen-Cooper-Schrieffer (BCS) superfluid to a Bose-Einstein Condensate (BEC),  and a non-interacting fully polarized Fermi gas for $x=0$. What happens for intermediate values of the asymmetry $x$ is  still a  matter of debate. In the case of a unitary Fermi gas strong theoretical arguments \cite{bf} point to the existence of an exotic new phase, a supersolid, with a pairing mechanism conjectured more than forty years ago \cite{fflo}, for asymmetries $0.25<x<0.75$. For larger asymmetries (smaller $x$) many authors have argued that the system is a normal Fermi system \cite{sandro}, a fact which has not been reconciled yet with the Kohn-Luttinger conclusions \cite{kohn}. 
The theoretical arguments are based largely on biased Monte-Carlo calculations, which by design were not able to reveal the existence of other phases. At the same time the experiments \cite{mit} were either insensitive or not designed to reveal the presence of other phases. Therefore the question of superfluidity of unitary asymmetric Fermi systems at very large asymmetries (small $x$) is still an open question, both theoretically and experimentally. 

Theory shows that away from unitarity for large asymmetries and zero temperature a fermion system is superfluid \cite{bfs}. In particular, on the BEC side of the Feshbach resonance, the ground state of a system is a mixture of two coexisting superfluids: a bosonic superfluid made of $s$-wave dimers and a P-wave fermionic superfluid  made out of the surplus majority fermions. These surplus fermions, which all have the same spin, interact with each other only through the density fluctuations of the bosonic superfluid in the $p$-wave or higher odd orbital states.  
P-wave fermionic superfluidity has been put in evidence so far in one system only, liquid $^3\text{He}$ 
\cite{helium}, but it is expected to appear in many other systems as well: among neutrons in neutron stars at densities slightly above nuclear densities \cite{lombardo}, in asymmetric cold gases with S-wave zero-range interaction \cite{bfs,bbf}, and in electronic systems \cite{electrons}. There are many studies suggesting the creation of P-wave superfluidity in cold fermionic gases \cite{iskin} using a P-wave Feshbach resonance \cite{jin} (instead of a S-wave Feshbach resonance), which is a different mechanism from the one considered by us here. The shorter lifetime of P-wave Feshbach molecules could prove to be a difficult experimental challenge to overcome in order to create these types of superfluid systems. On the other hand the S-wave Feshbach molecules are rather long lived and are routinely created in many laboratories \cite{labs}. 
The arguments for the existence of P-wave superfluids in our system based on the BCS formalism in the weak coupling limit was presented in Refs. \cite{bfs,bbf} : When all the minority atoms  bind into dimers (for positive scattering lengths) and form a BEC condensate at zero temperature, the surplus majority atoms will interact with each other only through the exchange of Bogoliubov sound modes of the BEC dimer condensate and their pairing is allowed only in a P-wave or other odd partial waves, due to the exclusion principle. 
In this mechanism, momentum and energy exchanged between the fermions can be comparable or large when compared with the Fermi momentum and energy of the surplus fermions. For this reason we expect that neither the BCS approximation, nor the more involved Eliashberg approximation with only energy dependence \cite{eliashberg}, in which only the retardation effects were accounted for, will be sufficient to describe quantitatively this kind of P-wave pairing mechanism, even though the calculation based on the BCS approximation in Ref \cite{bfs} is expected to be qualitatively correct. 

This paper is organized as follows : In section II, we describe the theoretical framework and give a brief description of our numerical calculation. In Section III, our numerical results on the induced P-wave pairing gap in homogeneous systems are discussed . In section IV, we discuss the induced P-wave pairing correlations in atomic traps and summarize our findings.

\subsection{II. Induced interaction and self-consistent equations} 

The mechanism for P-wave superfluidity in asymmetric Fermi gases is very similar to the BCS mechanism of superconductivity for electrons.  In an asymmetric Fermi gas on the BEC side of the Feshbach resonance (where the scattering length $a>0$) all the minority fermions form $N_b=N_\downarrow$ bosonic dimers with a ground state energy $\varepsilon_b=-\hbar^2/ma^2$. The surplus $N_f=N_\uparrow-N_\downarrow$ fermions (all with spin up in our convention) have the same spin and they interact with each other only via the excitation of the Bogoliubov sound modes in the BEC condensate of dimers, with which they coexist \cite{bfs}. The form of this interaction has been derived a long time ago by Bardeen, Baym and Pines \cite{bbp} and in the momentum-energy representation has the form 
\bea
\label{eq:ind}
U_{ind} (q,\omega) = \frac{2 U_{bf}^2n_b\varepsilon_q}{\omega^2 - 
   \varepsilon_q (\varepsilon_q + 2 n_b U_{bb})},  
\eea
where $q$ and $\omega$ are the momentum and energy exchanged between the two interacting fermions. 
$U_{bb}$ and $U_{bf}$ are the boson-boson and the boson-fermion couplings, respectively, 
$n_b=N_\downarrow/V$ is the boson (dimer) number density and $\varepsilon_q = q^2/(2 m_b)$ is the kinetic energy of a dimer with mass $m_b=2m$. $U_{bb} \approx 1.2\pi a /m $ and $U_{fb} \approx 3.54 \pi a /m$ are obtained from the fermion-fermion scattering length $a>0$ using few-body theory techniques, see Ref. 
\cite{bfs} for details.  In the usual BCS approximation one neglects the energy dependence of this induced interaction and then $U_{ind}$ becomes a Yukawa type exchange potential with a radius given by the coherence length of the BEC of dimers. The form of this induced interaction is generic for the coupling of fermions with sound modes.  Since the fermions couple to density variations each vertex is proportional to a gradient (and thus contributes ${\bm q}$ in momentum representation) and a coupling constant, and the propagating sound wave is described by the denominator in Eq. (\ref{eq:ind}). While we expect that for large values of the scattering length the coupling strength $U_{fb}$ could change from its weak coupling limit, the dependence of induced interaction $U_{ind} (q,\omega)$ on the energy and momentum should remain qualitatively unchanged.

It was Eliashberg \cite{eliashberg} who showed that retardation effects are very important. Depending on the magnitude of $\omega$ the induced interaction (\ref{eq:ind}) can change its character from attraction to repulsion. Tolmachev has shown a long time ago that the role of repulsion is somewhat diminished in the gap equation, see Ref. \cite{gennes}. In a more modern language that is translated into the fact that  the role of an attractive interaction is enhanced and a repulsive interaction is suppressed by the renormalization group flow.  The derivation of the Eliashberg gap equations \cite{eliashberg} relied on the famous Migdal's theorem \cite{migdal}, that the vertex corrections are small in the case of electrons. The role of the vertex corrections has been analyzed later and found to be important in the case of high $T_c$ superconductors \cite{vertex}, and one can expect that these corrections are also important in the case of dilute gases. 

In the simple Eliashberg approximation the pairing field has only an energy dependence.  In dilute Fermi gases however, as we will demonstrate here, both energy and momentum dependence of the induced interaction (\ref{eq:ind}) are crucial, as we will show by comparing the solutions of the simple Eliashberg equations and of the Dyson equations for the fermion self-energy with full momentum and energy dependence. The full energy-momentum dependence of the pairing gap equations has been investigated in condensed matter, nuclear physics and color superconductivity, see Refs. \cite{reuter}. 

As our results show, the pairing gap values obtained by taking into account the full energy-momentum dependence are enhanced considerably, especially towards the Feshbach resonance. Thus the prospects of putting in evidence for the first time this new type of superfluidity, when a bosonic superfluid coexists and leads to the formation of a fermionic P-wave superfluid, are greatly increased.

Here we solve the Dyson equation for the full propagator $G(p)$
\bea
\label{eq:G}
G^{-1}(p) =  [\omega - \varepsilon(p) \tau_3 -i \eta \omega ] - \Sigma (p),
\eea
where $p=(\omega,{\bm p})$ is the fermion 4-energy-momentum vector,
$\varepsilon(p)={\bm p}^2/2m-\mu$, $\mu$ the chemical potential, $\eta$ an infinitesimal small positive constant, and $\Sigma(p)$ is the self-energy. In this case the self-energy has the form
\bea
\label{eq:self}
\Sigma (p) &=& [1-Z(p)]\omega + \chi(p) \tau_3 + \phi(p) \tau_1, 
\eea
where $Z(p)$, $\chi(p)$, and $\phi(p)$ are the wavefunction renormalization, the Hartree-Fock potential, and the pairing field, respectively, and $\tau_{1,3}$ are Pauli matrices. From these equations one obtains the propagator $G(p)$
\bea
\label{eq:Gp}
G (p) = \frac{Z(p) \omega + \bar{\varepsilon}(p) \tau_3 + \phi(p) \tau_1}{[Z(p)\omega]^2 - E^2(p) + i \eta}, 
\eea
where $\bar{\varepsilon}(p)=\varepsilon(p)+\chi(p)$ and $E(p)=\sqrt{\bar{\varepsilon}^2(p)+|\phi(p)|^2}$.
The self-consistency is achieved through the equation for the self-energy 
\bea 
\label{eq:sigma}
\Sigma (p) =  i\int \frac{d^4q}{(2\pi)^4}U_{ind}(p-q)\tau_3G(q)\tau_3.
\eea
We will solve these equations by performing first a Wick rotation $\omega \rightarrow i\omega$, which will make all the integrands well behaved. One can show that this Wick rotation is equivalent to taking the zero temperature limit of the finite temperature Matsubara equations.  

Below we will compare the solutions for the pairing gap by solving both the simple Eliashberg equations and the Eqs. (\ref{eq:self}-\ref{eq:sigma}) for the self-energy retaining the full energy-momentum dependence. We will make {\it ans\"{a}tze} $Z(p)\propto Y_{00}({\hat{\bm p}})$, 
$\chi(p)\propto Y_{00}(\hat{\bm p})$ and $\phi(p) \propto Y_{1m}(\hat{\bm p})$, where $Y_{lm}(\hat{\bm p})$ are spherical harmonics of the unit vector $\hat{\bm p}={\bm p}/|{\bm p}|$. Also we will limit the magnitude of the exchanged energies $|\omega|\le \omega_D= c\hbar k_D$ and momenta $|{\bm p}|\le \hbar k_D$, where the Debye wave vector is defined by the average dimer-dimer (inverse) separation $k_D=\pi n_b^{1/3}$ and $c^2=n_bU_{bb}/m_b$ is the speed of the Bogoliubov sound modes. The physical pairing gap is given by $\Delta(p)=\phi(p)/Z(p)$, evaluated at $\omega=0$ and $|{\bm p}|=\hbar k_f$. The pairing gap $\Delta(p)=\Delta_{1m}Y_{1m}(\hat{\bm p})$ will be given in units of the Fermi energy of the surplus fermions $\varepsilon_f=\hbar^2k_f^2/(2m)$, where $n_f=(N_\uparrow-N_\downarrow)/V=k_f^3/(6\pi^2)$. 
The effect of the wavefunction renormalization is a change in the density of states and a reduction of the pairing gap. The fact that the equation for the self-energy is of convolution type allows us to use Fast Fourier Transform (FFT) techniques, after performing the integration over the angles analytically, and solve the self-consistent equations by iterations. When determining the self-consistent solution of Eqs. (\ref{eq:self}-\ref{eq:sigma}), the value of the chemical potential was fixed so as to have a prescribed ratio $n_b/n_f$. 

\begin{figure}[ht]
\includegraphics[width=9.0cm]{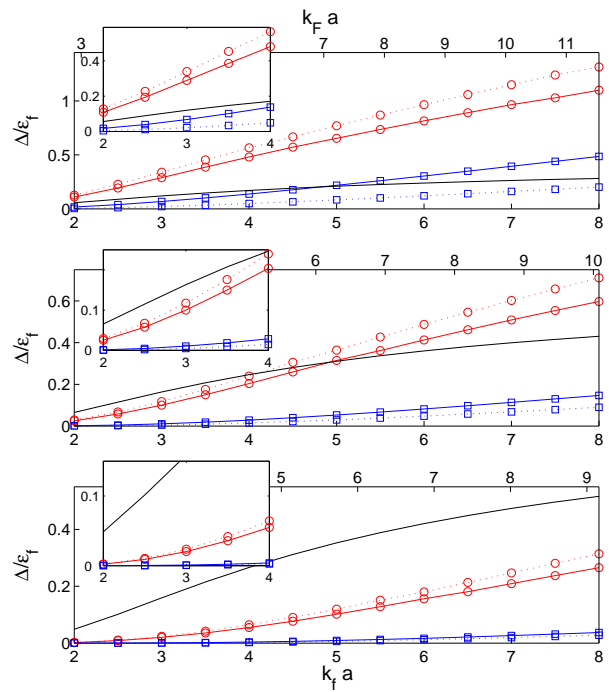}
\caption{\label{fig:gaps} (Color online) In these panels the ratio of dimer {\it versus} the surplus fermions varies as $y=n_b/n_f=2,1,$ and 0.5 from the upper panel down. The solid black line is the pairing gap computed using the BCS approximation as in Ref. \cite{bfs}. The solid (red) and dotted (red) line with circles are the pairing gaps computed with full momentum and energy dependence, while the solid (blue) and dotted (blue) line with squares are pairing gaps obtained in the simple Eliashberg approximation. The solid line is for the pairing gap $\Delta_{10}$ with $(l,m)=(1,0)$, while the dashed line is for $\Delta_{11}$ with $(l,m)=(1,1)$. Note that in the Eliashberg approximation $\Delta_{11}>\Delta_{10}$, while in the case of the BCS approximation there is no dependence of the pairing gap on the magnetic quantum number. The pairing gap is evaluated at zero energy $\omega =0$ and on the Fermi surface $|{\bm p}|=\hbar k_f$. In all cases we plotted the magnitude of the pairing gap, after the angular dependence $Y_{1m}(\hat{\bm p})$ has been factored out. The region of smaller values of $k_fa$ is shown in inserts.}
\end{figure}

The properties of this system are controlled by two dimensionless parameters, one the dimensionless strength of the induced interaction 
\bea 
\label{eq:lambda}
\lambda = \frac{mk_fn_b U_{fb}^2}{2\pi^2\varepsilon_{F}} \sim \frac{n_b}{n_f}(k_f a)^2 , 
\eea
and the other the ratio of the speed of sound to the Fermi velocity 
\bea 
\label{eq:s}
s = \frac{2 n_b U_{bb}}{\varepsilon_f} \sim \frac{c^2}{v_f^2} \sim \frac{n_b}{n_f}(k_f a). 
\eea
The asymmetry parameters $x$ and $y=n_b/n_f$ are related $y=x/(1-x)$. In defining the strength of the interaction one usually uses a different parameter $k_Fa$, where $k_F^3/(6\pi^2)=N_\uparrow/V$, and thus the relation between the two Fermi momenta is  $k_f =k_F(1-x)^{1/3}$ .

\subsection{III. Induced P-wave pairing gaps in homogeneous systems } 

In Fig. \ref{fig:gaps} we show the P-wave pairing gaps computed using various approximations, namely the pairing gap computed in the BCS approximation used in Refs. \cite{bfs,bbf}, the pairing gap computed in the simple Eliashberg approximation \cite{eliashberg}, and finally the pairing gap computed using the full momentum and energy dependence as described above. In the case of P-wave pairing gap one has to distinguish between two possible angular quantum numbers, namely $(l,m)=(1,0)$ when $\Delta_{10}(p)\propto p_z$ and the pairing gap has a nodal plane, and $(l,m)=(1,1)$ when $\Delta_{11}(p)\propto p_x+ip_y$ and the time reversal invariance is broken. In the BCS approximation the pairing gap has the same value in both cases. That is not the case either in the simple Eliashberg approximation or when the full energy-momentum dependence of the interaction is taken into account, and it appears that when full energy-momentum dependence is taken into account  $\Delta_{11} >\Delta_{10}$, unlike the case of the  simple Eliashberg approximation. 

The BCS approximation proves to be a rather inaccurate theory. The simple Eliashberg approximation underestimates the magnitude of the pairing gap by a significant factor, sometimes close to an order of magnitude. In hindsight all this does not come as a surprise, since for these systems, both magnitudes of the exchanged energy and momentum between two interacting fermions are large, and thus both effects due to a large range of the interaction and to retardation are naturally large.  It is also natural to find that the magnitude of the pairing gap is a monotonous function of the relative dimer density $y=n_b/n_f$. The dependence of the pairing gap on the scattering length $a$ is opposite to the dependence of the S-wave pairing gap in the symmetric phase. On the BEC side of the Feshbach resonance the pairing gap increases towards the BEC limit and tends to half of the binding energy of the dimer. The P-wave gap however shows the opposite behavior, it decreases in magnitude going from the Feshbach resonance towards the BEC limit (in which case the scattering length $a$ decreases).  The fact that the S-wave and the P-wave pairing gap have an opposite behavior is noteworthy. Another very important feature, which has already been noticed in Ref. \cite{bfs}, is that the magnitude of the pairing gap becomes comparable with the Fermi energy of the surplus fermions when approaching the Feshbach resonance. Our calculation does not include the vertex corrections however, which, as many authors have argued \cite{reuter}, are not expected to result in major changes.  

\subsection{IV. Aspects of induced P-wave pairing in atomic traps and final remarks} 

Since the induced P-wave pairing gap can be parametrically large while approaching the Feshbach resonance, the prospects to put it in evidence experimentally appear good. However, in current realizations of such systems in atomic traps, the role of the confining potential is going to be significant, and the realization of a homogeneous phase is unlikely. In a 3-dimensional trap one expects in the local density approximation a shell structure for an asymmetric system. The center of the trap will  be occupied by a pure paired unpolarized phase, followed by a mantle where both dimers and surplus atoms will coexist, and at the outskirts a pure unpolarized phase \cite{mantle}. It was shown in Ref. \cite{mantle} that in the weak coupling limit (when $a\rightarrow +0$) and in a harmonic trapping potential the radii of these shells are approximately related
\beq
6R^2_2 = R^2_3 + 5R^2_1,
\eeq
where $R_1$ is the radius of the pure dimer/unpolarized phase ($x(r\le R_1)=1$ and $n_f(r\le R_1)=0$), $R_2$ is the radius where the dimer number density vanishes ($n_b(r\ge R_2)=0$ and $x(r\ge R_2)=0$), 
and $R_3$ is the radius of the cloud ($n_f(R_3)=0$), thus $R_1<R_2<R_3$. 
Dimer and surplus majority atoms coexist within the shell $R_1<r<R_2$.  One can expect the appearance of a P-wave pairing only in this shell $R_1<r<R_2$ where the ratio $y=n_b/n_f$ changes from $y(R_1)=\infty$ to $y(R_2)=0$, while $\varepsilon_f(R_1)=0$ and has a finite value at $R_2$. Thus, assuming that the local density approximation is valid, we see that there are two competing factors in determining the magnitude of the P-wave pairing gap. The local Fermi energy $\varepsilon_f(r)$, which determines the order of magnitude of the P-wave pairing gap, increase from zero at $R_1$ to a final value at $R_2$
\beq
\varepsilon_f(r)\propto 5r^2+R^2_3-6R^2_2,
\eeq
while the dimer to fermion ratio has an opposite behavior, decreasing from infinity to zero
\beq
y(r)\propto \frac{R^2_2-r^2}{(5r^2+R^2_3-6R^2_2)^{3/2}},
\eeq
and consequently the P-wave pairing gap will attain a maximum absolute value in the middle of this shell. It will be a significant challenge for experimentalists to create a trap with a spatially rather large shell 
$R_1<r<R_2$ and a significant amount of surplus fermions so as to make visible this new exotic and quite unique in its nature P-wave paired phase.  It is important to remember that $k_fa=k_Fa(1-x)^{1/3}=
k_Fa/(1+y)^{1/3}<k_Fa$, where in a trap $k_f$ and $k_F$ are the local values. When one characterizes the proximity to the Feshbach resonance of a given Fermi gas, one typically uses the local Fermi momentum for the majority component at the center of the trap, which is significantly larger  than the local $k_F$ 
(and $k_f$) in the region $R_1<r<R_2$ of interest here. We expect that for finite values of the scattering length the radii $R_{1,2,3}$ will change, but these changes will not be qualitative and the conclusions drawn by us above about the structure of the mixed atom-dimer shell $R_1<r<R_2$ remain qualitatively correct.

In conclusion, we have determined the induced P-wave paring gap in an unbalanced system of atoms, where dimers coexist with atoms. 
Since the interaction between the like fermions is mediated by sound modes and it has a strong energy and momentum dependence, it could be either attractive or repulsive in character, depending on the particular values of the exchanged energy and momentum between the two atoms. We have compared the BCS induced P-wave pairing gap with pairing gap determined in the simple Eliashberg approximation, when only the energy dependence of the interaction is taken into account, and in the full Eliashberg approximation, when both momentum and energy dependence of the induced interaction is accounted for.  We have observed that both the BCS approximation and the simple Eliashberg approximation, with only a frequency dependence taken into account, do not lead to correct values of the pairing gap.  
Moreover, we have observed that by approaching the Feshbach resonance the pairing gap  
increases substantially and thus the conditions that induced P-wave pairing could be observed in atomic trap experiments improve significantly.  

The financial support through the US Department of Energy Grant No. DE-FG02-97ER41014 is gratefully acknowledged and we thank M.M. Forbes for a critical reading of the manuscript.  


\end{document}